\documentclass{article}

\begin{document}

\author{Bart D'Hooghe\thanks{Postdoctoral Fellow of the Fund for Scientific Research - Flanders
(Belgium)(F.W.O. - Vlaanderen).} \\
%EndAName
Clea, Vrije Universiteit Brussel (VUB)\\
Krijgskundestraat 33, 1160 Brussel, Belgium\\
\emph{e-mail: bdhooghe@vub.ac.be}}
\title{Communication through measurements and unitary transformations}
\date{}
\maketitle

\begin{abstract}
It is generally accepted that no `faster than light signalling' (FTLS) using
two entangled spin 1/2 particles is possible because of indeterminism in a
quantum measurement and linearity of standard quantum mechanics. We show how
in principle one bit of information could be transmitted using local
measurements and a global unitary transformation of the state of two
entangled spatially separated spin 1/2 particles. Assuming that the
postulate of a state collapse due to measurement is valid, the no FTLS
condition is saved if we do not have physical access to the required global
unitary transformation. This means that the no FTLS condition is also
present on the operational level, namely as imposing a physical restriction
on the possible realizable unitary transformations, in this case of two
entangled but spatially separated spin 1/2.
\end{abstract}

\section{Introduction}

The implications of an instantaneous collapse of the state of a quantum
system of two entangled particles due to a measurement has been a matter of
debate from the very beginning of quantum mechanics \cite{EPR, Bohm1951,
bell1964,aspect81,aspect82}. If two spin 1/2 particles are in a singlet
state, a local measurement on one of the particles provokes an immediate
change of the state (of the composite system and therefore also the state)
of the other particle, even if the two particles are spatially separated.
However, no `faster than light signalling' (FTLS) using local measurements
on a pair of two entangled spin 1/2 particles is possible because one does
not control the collapse of the spin. Gisin \cite{Gisin1989, Gisin1990},
reacting on the proposal of a nonlinear quantum mechanics by Weinberg \cite
{Weinberg1989a,Weinberg1989b}, put forward a clever way of sending signals
by a system of entangled spins in the case a nonlinear evolution would be
available, by using the entanglement of the reduced density states and by
coding one bit of information in the chosen measurement direction. In linear
quantum theory, this is not possible because the reduced density states are
independent of the chosen measurement direction such that the receiver
cannot decide by a local measurement which choice the sender has made.
Therefore, only in non linear modifications of quantum theory superluminal
signalling could be possible. This suggests that the no FTLS condition could
be used as a physical motivation for the linear structure of quantum
mechanics \cite{Gisin2001, Svetlichny98, fuchs2001, fuchs2002, clifbubhalv02}%
. However, non linear generalizations of quantum theory are possible in
which the no FTLS condition does hold \cite{Kent2002}. Also, choosing a
suitable nonlinear gauge transformation \cite{doebner} one can always
`disguise' a linear evolution equation into a non linear one. Therefore, non
linearity is identified as a necessary but not a sufficient condition for
FTLS.

In this paper we present a thought experiment which applies standard
(linear) quantum mechanics on a system of two entangled spin 1/2, with a
state evolution described by unitary transformations, and an instantaneous
collapse of the state if a measurement is performed. The thought experiment
uses two unitary transformations which are well-known in the theory of
quantum computation, namely the Hadamard gate and the Controlled NOT gate.
Hence it is natural to present our thought experiment with the concepts of
quantum computation (e.g. \cite{NielsenChuang}) and to talk about `qubits'
rather than `spin 1/2 particles'. First, we briefly recall some basic
properties of the Hadamard and the Controlled NOT gate to keep this paper
self-contained. Next, we show how to transmit the value of a bit by
performing local measurements and a unitary transformation (Controlled NOT)
on a system of two entangled (but possibly spatially separated) qubits.

\section{Signaling via CNOT and local measurements}

\subsection{Some unitary transformations used in quantum computation}

The Hadamard gate $H$ is a unitary operation acting on a single qubit,
mapping the vector $\left| 0\right\rangle ,$ respectively $\left|
1\right\rangle $, into the superposition vector $H\left| 0\right\rangle =%
\frac{\left| 0\right\rangle +\left| 1\right\rangle }{\sqrt{2}},$
respectively $H\left| 1\right\rangle =\frac{\left| 0\right\rangle -\left|
1\right\rangle }{\sqrt{2}}.$ The Controlled NOT gate $CNOT$ is a unitary
operation acting on two qubits mapping $\left| i,j\right\rangle $ onto $%
CNOT\left| i,j\right\rangle =\left| i,j\oplus i\right\rangle $, with $\oplus 
$ addition modulo 2. An intuitive view of this gate is that the value $j$ of
the `target' (second) bit is changed into its inverse $j\oplus 1$ whenever
the `control' bit $i$ has value $1$, and is left unchanged if the control
bit $i$ has value $0$. Nevertheless, one should keep in mind that this is
only an intuitive view of a unitary operation which acts on the two qubit
system as a whole. Indeed, in the basis $\left\{ \left| 0^{\prime
}\right\rangle =\frac{\left| 0\right\rangle +\left| 1\right\rangle }{\sqrt{2}%
},\left| 1^{\prime }\right\rangle =\frac{\left| 0\right\rangle -\left|
1\right\rangle }{\sqrt{2}}\right\} $ the roles of `control' and `target'
qubit are switched, such that $CNOT\left| i^{\prime },j^{\prime
}\right\rangle =\left| i^{\prime }\oplus j^{\prime },j^{\prime
}\right\rangle $; e.g. $CNOT\left| 0^{\prime },1^{\prime }\right\rangle
=\left| 1^{\prime },1^{\prime }\right\rangle $ and so on, showing that only
for a fixed choice of basis we can interpret the action of the CNOT gate in
terms of a control and a target qubit. Finally, we remark that $H=H^{-1}$
and $CNOT=CNOT^{-1}$.

\subsection{FTLS thought experiment}

Let the system of two qubits be prepared in the ground state $\left|
00\right\rangle .$ After a Hadamard transformation is applied to the first
qubit, a $CNOT$ is applied with the first qubit as `control' bit and the
second qubit as `target' bit. The state of the two entangled qubits is given
by $\psi _A$: 
\begin{eqnarray*}
\psi _A &=&CNOT\left( \left( H\otimes 1\right) \left| 00\right\rangle \right)
\\
&=&CNOT\left( \frac{\left| 00\right\rangle +\left| 10\right\rangle }{\sqrt{2}%
}\right) \\
&=&\frac{\left| 00\right\rangle +\left| 11\right\rangle }{\sqrt{2}}
\end{eqnarray*}
which is a maximally entangled state. Next, we assume that the two qubits
are spatially separated but stay in the entangled spin state $\psi _A=\frac{%
\left| 0_B0_A\right\rangle +\left| 1_B1_A\right\rangle }{\sqrt{2}}$ such
that Alice, the sender, has access to the second qubit, and Bob, the
receiver, has access to the first qubit. If Alice wants to send a bit value $%
1$ she performs a spin measurement in the computational basis $\left\{
\left| 0_A\right\rangle ,\left| 1_A\right\rangle \right\} $. If Alice wants
to send a bit value $0$ she does no measurement at all (hence not provoking
a state collapse). Let us denote the state after Alice has made her choice
by $\psi _{A^{\prime }}.$ Next, a `restoring procedure' is established by
applying $CNOT^{-1}=CNOT$ and $H^{-1}\otimes 1=H\otimes 1$ on the two qubits
system (we assume that this is possible, and discuss its validity and
consequences in next section). This state we denote by $\psi _B.$ Finally,
Bob who has access to the second qubit performs a spin measurement in his
computational basis $\left\{ \left| 0_B\right\rangle ,\left|
1_B\right\rangle \right\} .$ There are three possible events:

1) Alice does not perform a measurement (sending a bit value 0) and the
`restoring procedure' maps the state $\psi _{A^{\prime }}=\psi _A$ back into
the original initial state $\psi _B=\left| 00\right\rangle $: 
\begin{eqnarray*}
\psi _B &=&\left( H\otimes 1\right) \left( CNOT\left( \frac{\left|
00\right\rangle +\left| 11\right\rangle }{\sqrt{2}}\right) \right) \\
&=&\left( H\otimes 1\right) \frac{\left| 00\right\rangle +\left|
10\right\rangle }{\sqrt{2}}=\left| 00\right\rangle
\end{eqnarray*}
such that Bob obtains with certainty the outcome `$0$' in his (local)
measurement and the outcome $1$ has zero probability to occur.

2) Alice does perform a measurement (sending a bit value 1) and has observed
the outcome $1$. The state $\psi _A$ of the two entangled qubits has
collapsed in the state $\psi _{A^{\prime }}=\left| 11\right\rangle .$ After
applying the `restoring procedure', the state $\psi _B$ prior to the
measurement by Bob is given by: 
\begin{eqnarray*}
\psi _B &=&\left( H\otimes 1\right) CNOT\left( \left| 11\right\rangle \right)
\\
&=&\left( H\otimes 1\right) \left| 10\right\rangle \\
&=&\frac{\left| 00\right\rangle -\left| 10\right\rangle }{\sqrt{2}}
\end{eqnarray*}
such that Bob observes the outcome $0$ with probability 1/2 or the outcome $%
1 $ with probability 1/2.

3) Alice does perform a measurement (sending a bit value 1) and has observed
the outcome $0$. The state $\psi _A$ of the two entangled qubits has
collapsed in the state $\psi _{A^{\prime }}=\left| 00\right\rangle .$ After
applying the `restoring procedure', the state prior to the measurement by
Bob is given by: 
\begin{eqnarray*}
\psi _B &=&\left( H\otimes 1\right) CNOT\left( \left| 00\right\rangle \right)
\\
&=&\left( H\otimes 1\right) \left| 00\right\rangle \\
&=&\frac{\left| 00\right\rangle +\left| 10\right\rangle }{\sqrt{2}}
\end{eqnarray*}
such that again Bob observes the outcome $0$ with probability 1/2 or the
outcome $1$ with probability 1/2.

To conclude, if Bob observes an outcome $1$, he knows with certainty that
Alice wanted to send a bit value 1. If Bob observes the outcome $0$, Bob
cannot decide with certainty which of the three events has happened (i.e.,
whether Alice has performed a measurement or not). However, Alice and Bob
could use this procedure on a number $N$ of pairs of entangled qubits. E.g.,
for $N=10$, the probability that Alice performs a measurement (wanting to
send bit value 1) but Bob observes an outcome $0$ in each of his 10 spin
measurements, causing him to assume that actually a bit value $0$ was sent,
is $\left( 0.5\right) ^{10}\approx 0.1\%.$ Combining this redundancy
technique with classical bit correction techniques, this procedure could be
used to transmit bits of information with probability of Bob correctly
receiving the bit arbitrary close to unity.

\subsection{Discussion}

The thought experiment shows how it is in principle (\emph{mathematically})
possible using unitary transformations and state collapse due to measurement
to transmit one bit of information with probability of Bob correctly
receiving the bit arbitrary close to unity. Hence, if one wants to maintain
the no FTLS condition in \emph{physical} reality, one of the assumptions
made in the thought experiment has to be physically impossible. One
possibility is to drop the assumption that the collapse of the state due to
measurement is real. Another possibility is that not all unitary
transformations used in the thought experiment can be performed in reality.
Since the Hadamard gates are local unitary gates, working on a single qubit,
the only `impossible' unitary transformation should be the CNOT gate acting
on two spatially separated qubits. This means that the no FTLS condition is
not only determined by the linearity of standard quantum mechanics (equipped
with the state collapse postulate), but is also present on the operational
level, namely as imposing a physical restriction on the possible realizable
unitary transformations of a quantum system. It means that certain (which
appear to be mathematically in principle possible) unitary transformations
are physically impossible.

\section{Conclusions}

We have shown that in theory bits of information could be transmitted using
local measurements and a global unitary transformation on a system of two
entangled (but spatially separated) qubits, following the rules of standard
linear quantum mechanics with an instantaneous collapse of the state due to
measurement. Although it is impossible to control the collapse of the state
in a quantum measurement, whether a collapse has actually occurred or not
does make a difference. Therefore, the thought experiment shows that the no
FTLS condition is not just a consequence and possible physical justification
of the linearity of quantum mechanics, but also translates onto the
operational level as the impossibility to perform a certain global unitary
transformation (in this case the CNOT) on a pair of spatially separated
qubits.

\end{document}